# Deep Learning for Spin-Orbit Torque Characterizations with a Projected Vector Field Magnet

Chao-Chung Huang[1], Chia-Chin Tsai[1], Wei-Bang Liao[1], Tian-Yue Chen[1], and Chi-Feng Pai[1,2*]

[1]*Department of Materials Science and Engineering, National Taiwan University, Taipei 10617, Taiwan*

[2]*Center of Atomic Initiative for New Materials, National Taiwan University, Taipei 10617, Taiwan*

Spin-orbit torque characterizations on magnetic heterostructures with perpendicular anisotropy are demonstrated on a projected vector field magnet via hysteresis loop shift measurement and harmonic Hall measurement with planar Hall correction. Accurate magnetic field calibration of the vector magnet is realized with the help of deep learning models, which are able to capture the nonlinear behavior between the generated magnetic field and the currents applied to the magnet. The trained models can successfully predict the applied current combinations under the circumstances of magnetic field scans, angle scans, and hysteresis loop shift measurements. The validity of the models is further verified, complemented by the comparison of the spin-orbit torque characterization results obtained from the deep-learning-trained vector magnet system with those obtained from a conventional setup comprised of two separated electromagnets. The damping-like spin-orbit torque (DL-SOT) efficiencies ($|\xi_{DL}|$) extracted from the vector magnet and the traditional measurement configuration are consistent, where $|\xi_{DL}| \approx 0.22$ for amorphous W and $|\xi_{DL}| \approx 0.02$ for $\alpha$-W. Our work provides an advanced method to meticulously control a vector magnet and to conveniently perform various spin-orbit torque characterizations.

---

[*] Email: cfpai@ntu.edu.tw

# I. INTRODUCTION

Current-induced spin-orbit torques (SOTs) originating from the spin Hall effect (SHE) [1,2], orbital Hall effect [3-5] or interfacial Rashba-Edelstein effect (REE) [6] are able to manipulate magnetization of ferromagnetic layers and induce magnetization switching [6], microwave frequency oscillation [7,8] and domain wall motion [9-11], providing versatile approaches to realize SOT driven magnetic random access memory (MRAM) [12,13], nano-oscillator [14] and spin logic gates [15,16]. Due to the high potential of spintronics devices, SOTs of novel materials have been widely investigated with several characterization methods including spin torque ferromagnetic resonance (ST-FMR) [17], harmonic Hall technique [18,19] and hysteresis loop shift measurement [11]. However, it is hard to perform all these measurements within a compact setup since they require different combinations of magnetic field. For instance, ST-FMR and harmonic Hall for samples with perpendicular magnetic anisotropy (PMA) [19] need in-plane magnetic fields with varying strengths along the same direction, while a sweeping out-of-plane field and a constant in-plane field are necessary for the hysteresis loop shift technique. Moreover, angular transport measurements like harmonic Hall for samples with easy-plane magnetic anisotropy [18], spin Hall magnetoresistance (SMR) [20,21], and planar Hall effect (PHE) require a rotating in-plane field with a fixed field magnitude. To perform these angle-dependent measurements, the measured devices are typically wire bonded and mounted on a rotary stage while an electromagnet provides magnetic field along a fixed direction. However, the wire bonding process could be time-consuming and limit the number of the samples being tested, making such angular transport measurements inconvenient.

In this paper, we show that a projected vector field magnet can be controlled via deep learning models with a single hidden layer composed of 9 and 18 nodes, which are able to link the current sets applied to the vector magnet to the output resultant field. The models can capture the complexity and the nonlinearity of the generated magnetic fields, allowing the magnet to output accurate fields for field scans ($x, y, z$), angle scans ($\theta, \varphi$) and hysteresis loop shift measurements. The feasibility of the models is systematically examined with detailed field measurements for each case. It is found that the measured magnetic fields are in line with the desired values, which clearly justifies the precision of the trained models and the versatility of the vector magnet. More importantly, SOT characterization is performed by means of the hysteresis loop shift measurement on the sputtered W/Co-Fe-B magnetic heterostructures with the trained vector magnet system. The damping-like SOT (DL-SOT) efficiency $\xi_{DL}$ of W/Co-Fe-B shows a trend of phase transition. When the W thickness is less than 5 nm, $|\xi_{DL}|$ has a maximum of ~ 0.22. As its thickness is greater than 5 nm, $|\xi_{DL}|$ reduces down to ~ 0.02. To further confirm the practicability of the loop shift protocol on the magnet, we repeat identical measurements with a traditional setup that consists of two sets of electromagnets. The consistency between the results obtained from the two setups again validates the training models. Finally, we demonstrate harmonic Hall measurement with PHE correction using the vector field magnet, where the angular scan is used for the small field approximation of planar Hall resistance and field scan is utilized for harmonic Hall measurement. The measurement protocols can be realized within the same setup, thereby showing the flexibility of the vector field magnet.

# II. MODEL TRAINING

The projected vector field magnet used in this work is GMW Associates Model 5204, which mainly consists of three triangular electromagnet poles arranged 120 degrees between each other. A schematic illustration of the vector magnet is shown in Fig. 1(a), where the three poles are labeled as 0, 1 and 2. Each pole is driven with a Kepco BOP10-100MG power supply independently. A National Instruments USB-connected data acquisition system (DAQ) serves as a medium to send commands of applied currents to the power supplies. An industrial chiller is used to keep the magnet coils working at a constant temperature of around 20°C, avoiding a resistance change of the coils due to heating. With different combinations of the currents applied to the three electromagnets, the vector magnet can overall produce a field along a given direction and magnitude. For example, to produce a magnetic field along $x$-axis, one should apply current to pole 0. As for the case of magnetic field along $y$-axis, one should apply opposite currents to pole 1 and 2 at the same time, as schematically shown in the intersection plots of the vector magnet in Fig. 1(b). Although a linear current-to-field response is typically expected, the magnetic field provided by an electromagnet could become nonlinear to large applied currents because of the saturation effect of an electromagnet pole that a sharp increase of the applied current is required after the saturation of iron core's magnetization. Also, the behavior of magnetic flux could also become complicated while different currents are applied to each triangular electromagnet pole simultaneously. In order to resolve these problems and perform trust-worthy SOT characterizations based on the vector magnet, two deep learning models are employed.

First of all, a database of measured magnetic field and applied current sets is established. A triaxial Hall sensor which can measure magnetic field along three orthogonal directions at the same time is used to collect the training dataset of magnetic field. The Hall sensor is connected to the DAQ system, allowing simultaneous readings of the three voltage values measured, which are proportional to the measured field components along the three directions. As illustrated in Fig. 1(c), the sensor is fixed with a probe holder and placed perpendicular to the surface of the vector magnet with its orientation consistent with the coordinate system in Fig. 1(a). Next, we apply around 300,000 current combinations and measure the corresponding magnetic fields with the triaxial Hall sensor. The obtained data is plotted as a field map shown in Fig. 2(a), which shows a clear three-fold symmetry due to the arrangement of the three electromagnet poles. According to the field map, the vector magnet can provide in-plane field toward any direction up to 2000 Oe with an out-of-plane component less than 50 Oe. Precise in-plane field can even be applied up to around 2500 Oe at specific directions. As for out-of-plane field, the limit is around 1600 Oe.

To further utilize the vector magnet and to capture the relation between applied currents and magnetic field, we utilize deep learning models to train the measured data. The dataset first undergoes preprocessing including random shuffling and standardization. After that, the dataset is split into a training set and a test set with ratio 4:1, respectively. The comparison between the learning curves of the two datasets can tell whether the data is overfitted or underfitted, providing us a path to tune the training parameters. The training then begins after the above preprocessing. Note that here the inputs of models are the measured magnetic fields and the outputs are the applied currents, since we expect the models to predict current combinations corresponding to the desired applied field. For the two deep learning models used in this work, both network structures are composed of one input layer, a hidden layer, and an output layer, as shown in Fig. 2(b). The hidden layer of the first model consists of 9 nodes (abbreviated as M9) and the second one has 18 nodes (abbreviated as M18). Hyperbolic tangent is chosen as the input activation function because our input and output both include positive and negative values. The output activation function is simply selected as the linear function as our output is beyond the range of hyperbolic tangent. Since the models are performing regression, the loss function is chosen as mean square error (mse). The adopted optimizer is the adaptive moment estimation algorithm (Adam) with learning rate ($\alpha$) and batch size set as 0.01 and 10000, respectively. As shown in Fig. 2(c), the learning curves (loss as a function of epochs) of the training set and the test set have almost identical trajectories during training and converge close to 0 after 300 epochs, indicating that the model M9 is well-trained. For the case of model M18, the parameters set for training are identical as above. The training results and learning curves are similar to the case of model M9. The details of the code for training can be referred to our GitHub repository [22].

## III. MODEL PREDICTIONS AND EXAMINATIONS

To look into how the trained models can grasp the nonlinear nature of the vector magnet, we examine the predictions of applied currents given by the models and scrutinize them with the triaxial Hall sensor under several magnetic field sets which are often used in transport measurements. Note that in this work, model M9 is adopted when out-of-plane field is involved and model M18 is mainly used when only in-plane field is applied, since we find that they have better performance under specific scenarios. The first field sets are magnetic field scans along three axes $x$, $y$ and $z$, which are the setups used frequently for standard transport measurements like anisotropic magnetoresistance (AMR) and anomalous Hall effect (AHE). The current predictions of field scan along $x$, $y$ and $z$-axis are shown in Fig. 3(a), (b) and (c), respectively. According to Fig. 3(a), pole 0 serves as the main contribution of magnetic field along $x$-axis ($H_x$). For the other two poles, small currents with sign opposite to that of pole 0 are applied to eliminate the redundant out-of-plane field. When the nominal applied $H_x$ is greater than 1500 Oe, the calculated applied currents begin to show nonlinearities. Applied current of pole 0 ($I_0$) increases exponentially while $I_1$ and $I_2$ slightly increase then decrease within the nonlinear regime. Next, the model predictions of field scan along $y$-axis are inspected. For the $y$-field scan case, pole 1 and 2 now become the dominant $H_y$ source. To provide pure $H_y$, opposite currents are applied to pole 1 and pole 2 and the nonlinearity of the predicted applied currents occur when the applied $H_y$ is greater than 2000 Oe. Within the current limit of vector magnet, the maximum of applied $H_y$ given by the deep learning models are 2500 Oe, which is consistent with the field map shown in Fig. 2(a). Unlike $x$ and $y$ field scans, the predictions of $z$ field scan seem to be much simpler. Nearly identical currents should be applied to the three electromagnet poles in order to get rid of the in-plane components of magnetic field, as shown in Fig. 3(c). The result seems quite intuitive owing to the symmetry of the pole arrangement. More importantly, the current predictions of field scans given by the models are further examined with real field measurements. The same triaxial Hall sensor is utilized to carry out the inspection. The measured fields with respect to the nominal applied fields of *x*, *y* and *z* field scan are shown in Fig. 3(d), (e) and (f), respectively. Small field results are also shown in Fig. 3(g), (h), and (i). The measured fields along the three axes are linear to the nominal applied fields with negligible field on the other two axes. The relative errors $\Delta H_i$ ($i = x, y, z$) for the three field scans are shown in the insets of Fig. 3(d), (e), and (f), respectively. Despite the rather sizable errors (20 to 50%) at small applied fields (< 100 Oe) which could be originated from the hysteresis of the electromagnet poles, the overall relative errors are small (~ 5%) for all three field scans, indicating that the deep learning models do serve as effective mediums to communicate with the vector magnet for field scan measurements.

Next, we study the applied current predictions of $\varphi$ and $\theta$ angle scans with a constant magnetic field magnitude. First, we check the $\varphi$ angle scan. The predicted applied currents with field 1000 Oe and 1900 Oe are shown in Fig. 4(a) and (b), respectively. For 1000 Oe, the applied currents of the three poles show sinusoidal trends with 120 degrees phase shift, which seems intuitive as 1000 Oe is still within the linear regime of *x* and *y* field scan. However, as the field is increased to 1900 Oe, the predicted currents show more complicated behavior. The nominal field angular dependence of applied currents are no longer simple sinusoidal functions. The observation is consistent with the nonlinearities observed in Fig. 3(a) and (b). In addition, we further check the predicted currents of $\theta$ angle scan at $\varphi = 0$ and $\varphi = 90°$, which are shown in Fig. 4(c) and (d) respectively. The applied currents of the three poles show cosine trends for both $\theta$ angle scan cases as a whole with some subtle differences. For the scenario of $\varphi = 0°$, $I_0$ is positive while the other two are negative when $\theta = 90°$. As for the case of $\varphi = 90°$, $I_1$ and $I_2$ have opposite signs while $I_0$ is negligible when $\theta = 90°$. The characteristics are again quite consistent with the schemes of *x* and *y* field scan. After the qualitative investigation of applied currents at various angle scan conditions, the validity of the predicted currents is again examined with real field measurements. The measured magnetic fields along three axes as a function of applied field angle are shown in Fig. 5(a) and (b) for $\varphi$ angle scan with in-plane field ($H_{in}$) 1000 Oe and 1900 Oe, respectively. In both cases, $H_x$ and $H_y$ can be well fitted by a cosine and a sine function with negligible $H_z$. We further plot the $\varphi$ angle scan field maps with various field magnitudes, which are shown in Fig. 5(c). The field contours are smooth at different field magnitudes. On top of that, as shown in Fig. 5(d), the measured field amplitude is linear to the nominal applied field, indicating that the currents derived from the deep learning models induce feasible magnetic field for $\varphi$ angle scan. Additionally, the models also give decent magnetic field under the $\theta$ angle scans on *xz* plane and *yz* plane, as shown in Fig. 6. In short, the deep learning models M9 and M18 are able to evaluate current sets applied to the vector magnet, which provides precise magnetic fields for both $\varphi$ and $\theta$ angle scans. Angular transport properties can therefore be probed directly on a vector field probe station without tedious wire bonding process.

Finally, we check the performance of the models for the magnetic field combinations used in hysteresis loop shift measurement. In the measurement protocol, a constant in-plane field $H_x$ is applied to compensate the Dzyaloshinskii-Moriya interaction effective field and an out-of-plane field $H_z$ is swept to probe the SOT-induced effective field [11]. The two perpendicular fields are conventionally generated from two separate sets of electromagnets. The current predictions of the magnetic field set with $H_x$ = 500 Oe and 1000 Oe are shown in Fig. 7(a) and (b), respectively. In general, the current combinations are quite similar to that for typical $z$ field scan shown in Fig. 3(c). The applied currents of the three poles are almost the same and increase linearly with nominal $H_z$. Due to the additional $H_x$, $I_0$ is slightly deviated from the linear trend and the deviation becomes larger as the applied $H_x$ increases. The scheme can be considered as the superposition of currents for $x$ and $z$ field scans in the linear regime. Again, we test the validity of the predicted currents with the Hall sensor. The measured magnetic field components as functions of nominal applied $H_z$ of $H_x$ = 500 Oe and 1000 Oe are shown in Fig. 7(c) and (d), respectively. The measured $H_x$ is indeed fixed at 500 Oe and 1000 Oe while $H_z$ is swept linearly from -600 to 600 Oe, which is in line with the field combination utilized for loop shift measurement.

As a result, the deep learning models well describe the intricate behaviors of the projected vector field magnet, including the nonlinearities between applied currents and output field. Thanks to the models, the projected vector field magnet allows us to perform simple field scans, angle scans, and loop shift measurement. However, the field limitations could be one of the shortcomings of the projected vector field magnet. For instance, the maximum $H_z$ allowed in the current setup is only 1600 Oe, which is insufficient to saturate many commonly-seen in-plane magnetized ferromagnets along out-of-plane direction. This could be tentatively solved by incorporating larger electromagnets. Field homogeneity is also a critical issue as the region with uniform magnetic field (no greater than 1 mm by 1 mm) is much smaller than typical electromagnets made up of split coils (depends on the gap size). Therefore, a precise positioning of the device under test would be challenging, which is typically mitigated by the adoption of a micropositioner. Nevertheless, the projected vector field magnet system complemented with a deep-learning-assisted calibration still provides decent and flexible magnetic fields under various circumstances, showing the potential to efficiently conduct a large-scale probing and the versatility to perform various kinds of SOT characterizations.

## IV. HYSTERESIS LOOP SHIFT MEASUREMENT

To field test on real devices, we perform hysteresis loop shift measurement on a series of W/Co-Fe-B magnetic heterostructures with PMA. W($t_W$)/Co$_{40}$Fe$_{40}$B$_{20}$(1.4)/MgO(2)/Ta(2) (numbers in the parenthesis are in nanometers) are deposited onto Si/SiO$_2$ substrate with high-vacuum magnetron sputtering (base pressure ~ $10^{-8}$ Torr), with $t_W$ ranges from 2 nm to 7 nm. The working Ar pressure is 3 mTorr (10 mTorr) for DC (RF) sputtering metallic (oxide) layers. The samples are post-annealed at 300°C for 1 hr to induce PMA. The Hall bar devices with lateral dimensions of 5×60 μm$^2$ are then fabricated from the sputtered films with photolithography and a subsequent patterning process.

In the protocol of hysteresis loop shift measurement, a DC current $I_{DC}$ is applied to the Hall bar device and the Hall resistance is measured under a constant $H_x$ while sweeping $H_z$. The measured Hall resistance shows a hysteresis loop due to the AHE. To counteract the interfacial Dzyaloshinskii-Moriya interaction effective field ($H_{DMI}$) and to induce domain wall propagation, $H_x$ is applied to align the domain wall moment that the current-induced SHE exerts an out-of-plane damping-like effective field ($H_{eff}^z$) on the moment, thereby generating the hysteresis loop shift phenomenon [9,11]. As the domain wall moment is fully aligned, the out-of-plane damping-like effective field per applied current ($H_{eff}^z/I_{DC}$) saturates with respect to $H_x$. The DL-SOT efficiency ($\xi_{DL}$) can then be quantified with the saturated $H^z_{eff}/I_{DC}$ according to the equation below [11,23]:

$$\xi_{DL} = \left(\frac{2}{\pi}\right)\frac{2e\mu_0 M_s t_{CoFeB} w t_W}{\hbar}\left(\frac{\rho_{CoFeB} t_W + \rho_W t_{CoFeB}}{\rho_{CoFeB} t_W}\right)\left(\frac{H^z_{eff}}{I_{DC}}\right), \qquad (1)$$

where $\mu_0$ is the vacuum permeability, $M_s$ is the saturation magnetization, $t_{CoFeB}$ is the thickness of Co-Fe-B layer, $w$ is the width of the Hall bar device, $\rho_{CoFeB} = 190\ \mu\Omega \cdot cm$ is the resistivity of CoFeB and $\rho_W$ is the resistivity of W, which depends on its phase. For amorphous-W, $\rho_W = 200\ \mu\Omega \cdot cm$, while for $\alpha$-W, $\rho_W = 30\ \mu\Omega \cdot cm$. Saturation magnetization of the Co-Fe-B layer is 1200 emu/cm$^3$, as characterized by a vibrating sample magnetometer (VSM) [24]. Here the hysteresis loop shift measurement is performed with (1) the deep-learning-trained vector magnet system and (2) the conventional setup including two electromagnets for independent $H_x$ and $H_z$. As shown in Fig. 8(a) and (b), the hysteresis loop shift phenomenon can be observed with the vector magnet and the extracted $H^z_{eff}$ shows a linear relation with the applied current, which is consistent with previous studies [9,11]. We then compare the hysteresis loop shift results obtained from the two setups. Fig. 8(c) shows $H^z_{eff}/I_{DC}$ as a function of $H_x$ for the W(5)/Co-Fe-B(1.4) sample obtained from two measurement configurations, from which similar trends can be observed. This consistency indicates that a well-trained vector magnet system is indeed practicable for performing hysteresis loop shift measurements.

The estimated DL-SOT efficiencies $|\xi_{DL}|$ of the W-based samples measured with the two setups are shown in Fig. 8(d), whose inset is the relative deviation between the two setups. The

uncertainties of the estimated $|\xi_{DL}|$ stem from the linear regression of $H_{eff}^z$ with respect to $I_{DC}$. Here, $|\xi_{DL}|$ reaches a maximum (~ 0.22) at $t_w$ = 3 nm and proceeds to decrease as $t_w$ increases further. At $t_w$ = 7 nm, $|\xi_{DL}|$ decreases to a minimum (~ 0.02). The trend is related to the phase transition of W from an amorphous phase to a polycrystalline $\alpha$-W phase, which is consistent with the results from several previous literatures [25-27]. The $\xi_{DL}$ values measured with the projected vector field magnet are fairly consistent with the results obtained with the conventional configuration that the maximum relative deviation is around 30% of the estimated value, again confirming that the vector magnet is applicable for hysteresis loop shift measurements.

## V. HARMONIC HALL MEASUREMENT

To further demonstrate the versatility of the projected field vector magnet, we perform harmonic Hall measurement on the W(5)/Co-Fe-B(1.4)/MgO(2)/Ta(2) Hall bar device. In this measurement, an AC current with $I_{AC}$ = 1.2 mA and frequency $f$ = 171 Hz is applied to the device while either longitudinal ($H_x$) or transverse field ($H_y$) is swept. The harmonic signals are captured with the standard lock-in technique. The first harmonic signals implicate the information of the magnetization equilibrium position and it is widely used for the extraction of the anisotropy field, while the current-induced damping-like ($H_{DL}$) and field-like ($H_{FL}$) SOT effective fields can be detected with the second harmonic signals. Fig. 9 (a) shows representative first harmonic results with field swept along $x$-axis. The solid curves are fitted according to the following equation [28]:

$$V_H^{1\omega} = V_{AHE}\left(1 - \frac{1}{2}\left(\frac{H_{x(y)}}{H_k}\right)^2\right), \quad (2)$$

where $V_{AHE}$ and $H_k$ are the anomalous Hall voltage and the anisotropy field, respectively. Here $H_k$ of the W(5)/Co-Fe-B(1.4) sample is estimated to be 5100 Oe with eqn. (2).

As for the second harmonic results, the Hall voltages are linearly proportional to the applied $H_x$ and $H_y$, as shown in Fig. 9(b) and (c). For different initialized states, the slopes are both positive when $H_x$ is swept, while the slopes have opposite signs in $H_y$ scan. The observed field dependence is consistent with previous harmonic Hall studies [29,30]. The current-induced DL-SOT and FL-SOT effective fields can then be quantified with the following equation [19]:

$$H_{DL(FL)} = -\frac{2}{\zeta} \cdot \frac{\beta_{L(T)} + 2p\beta_{T(L)}}{1 - 4p^2}, \quad (3)$$

where $\zeta = \partial^2 V_H^{1\omega}/\partial H_x^2$ and $\beta_{L(T)} = \partial V_H^{2\omega}/\partial H_{x(y)}$, which can be obtained with the first harmonic fittings following eqn. (2) and linear fittings of second harmonic, respectively. Here $p = R_{PHE}/R_{AHE} = 0.33$, which is the ratio between planar Hall and anomalous Hall resistance. The term is included since the PHE can contribute to the harmonic signals, thereby affecting the values of extracted effective fields. To estimate $R_{PHE}$, $\varphi$ angle scans with various small fields are performed with the aid of the trained models. A small angle approximation of $R_{PHE}$ is adopted in accordance with the following equation [31]:

$$R_H = R_{PHE} \left(\frac{H_{in}}{H_k}\right)^2 \sin2\varphi = R'_{PHE} \sin2\varphi, \tag{4}$$

where $R'_{PHE}$ is the fitted $\sin2\varphi$ amplitude. As shown in Fig. 9(d), the measured Hall resistances show typical $\sin2\varphi$ trends with respect to the applied field angle, which verifies the applicability of field for $\varphi$ angle scans. As further shown in Fig. 9(e), the fitted amplitude $R'_{PHE}$ linearly increases with $H_{in}^2$, which is in line with eqn. (4). With the linearly fitted slope and $H_k$ obtained previously from the first harmonic results, $R_{PHE}$ is approximated as 1.27 Ω. As all the parameters are collected from eqn. (2) to (4), the DL-SOT and FL-SOT efficiencies can then be estimated with the following equation [32,33]:

$$\xi_{DL(FL)} = \frac{2e\mu_0 M_s t_{CoFeB} w t_W}{\hbar} \left(\frac{\rho_{CoFeB} t_W + \rho_W t_{CoFeB}}{\rho_{CoFeB} t_W}\right)\left(\frac{H_{DL(FL)}}{I_{AC}}\right). \tag{5}$$

The resultant $|\xi_{DL}|$ and $|\xi_{FL}|$ of the W(5)/Co-Fe-B(1.4) sample are found to be ~ 0.38 and 0.056, respectively. Note that a recent work has proposed that PHE induced by SOT gives negligible contribution to second harmonic even when $R_{PHE}$ is sizable compared to $R_{AHE}$ [31]. Under the circumstance, the DL-SOT and FL-SOT effective field can be simply quantified with the following equation[19]:

$$H_{DL(FL)} = -\frac{2}{\zeta} \cdot \beta_{L(T)}. \tag{6}$$

Without the consideration of the PHE correction, $|\xi_{DL}|$ and $|\xi_{FL}|$ of the W(5)/Co-Fe-B(1.4) sample are estimated to be ~ 0.34 and 0.097, respectively.

In general, the DL-SOT efficiency obtained from harmonic approach is larger than that obtained from hysteresis loop shift measurement. The discrepancy between the two results is ascribed to the different assumptions between the two techniques. Hysteresis loop shift is based on a micromagnetic framework (multi-domain) while harmonic Hall protocol is established on a macrospin scheme (single domain). Moreover, thermal effects originated from the ordinary and anomalous Nernst effect can contribute to unwanted second harmonic signal as well [18,34], leading to possible overestimation of $|\xi_{DL}|$. In short, the PMA harmonic Hall protocol with PHE correction, which includes both field scan and angle scan measurements, is demonstrated on the compact vector field magnet system.

## VI. CONCLUSIONS

To conclude, the projected vector field electromagnet system can be well controlled to provide different magnetic field combinations (including field scans, angle scans and hysteresis loop shift measurement) with decent precision through deep learning models with a single hidden layer. At these circumstances, the relations between applied magnetic field and currents of the vector magnet derived with the models are carefully examined with detailed field measurements via a triaxial Hall sensor. The validity of the trained models are further verified by actual hysteresis loop shift measurement with the projected vector field magnet on a series of W-based samples with PMA. The obtained DL-SOT efficiencies are in line with the results from a traditional loop shift configuration. Furthermore, harmonic Hall measurement with PHE correction can be performed on the identical setup. The development of the deep learning models allows accurate application of arbitrary magnetic field on the projected vector field magnet. The compact versatile setup makes various SOT characterization approaches possible with fewer steps of fabrication or device preparation process.

## ACKNOWLEDGEMENTS

We thank Kuan-Hao Chen for the assistance on setting up the triaxial Hall sensor. This work is supported by the Ministry of Science and Technology of Taiwan (MOST) under grant No. 111-2636-M-002-012 and by the Center of Atomic Initiative for New Materials (AI-Mat), National Taiwan University from the Featured Areas Research Center Program within the framework of the Higher Education Sprout Project by the Ministry of Education (MOE) in Taiwan under grant No. NTU-110L9008.

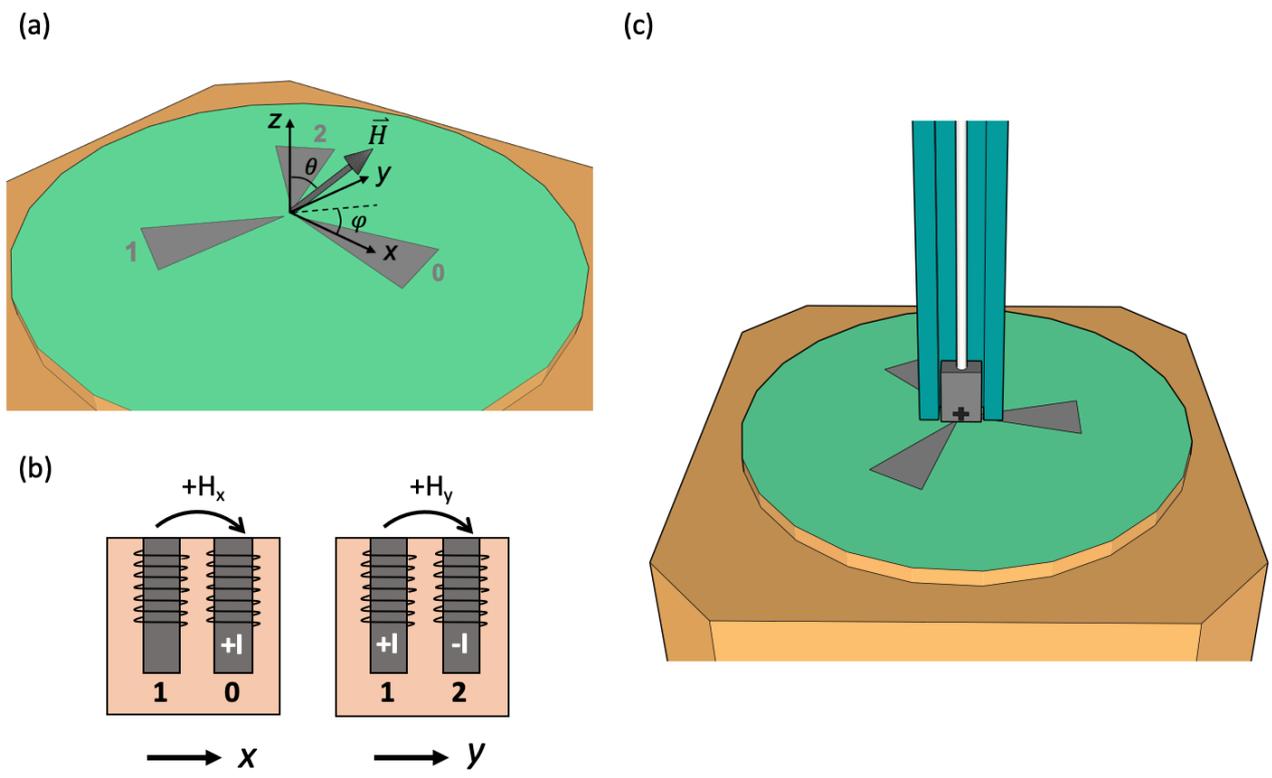

Figure 1. (a) Schematic illustration of the projected vector field magnet with the three poles labeled as 0, 1 and 2. (b) Schematic intersection plots of the projected vector field magnet along $x$ and $y$-axis. (c) Schematic plot of the triaxial Hall sensor setup. The sensor is fixed by a probe holder during magnetic field measurements.

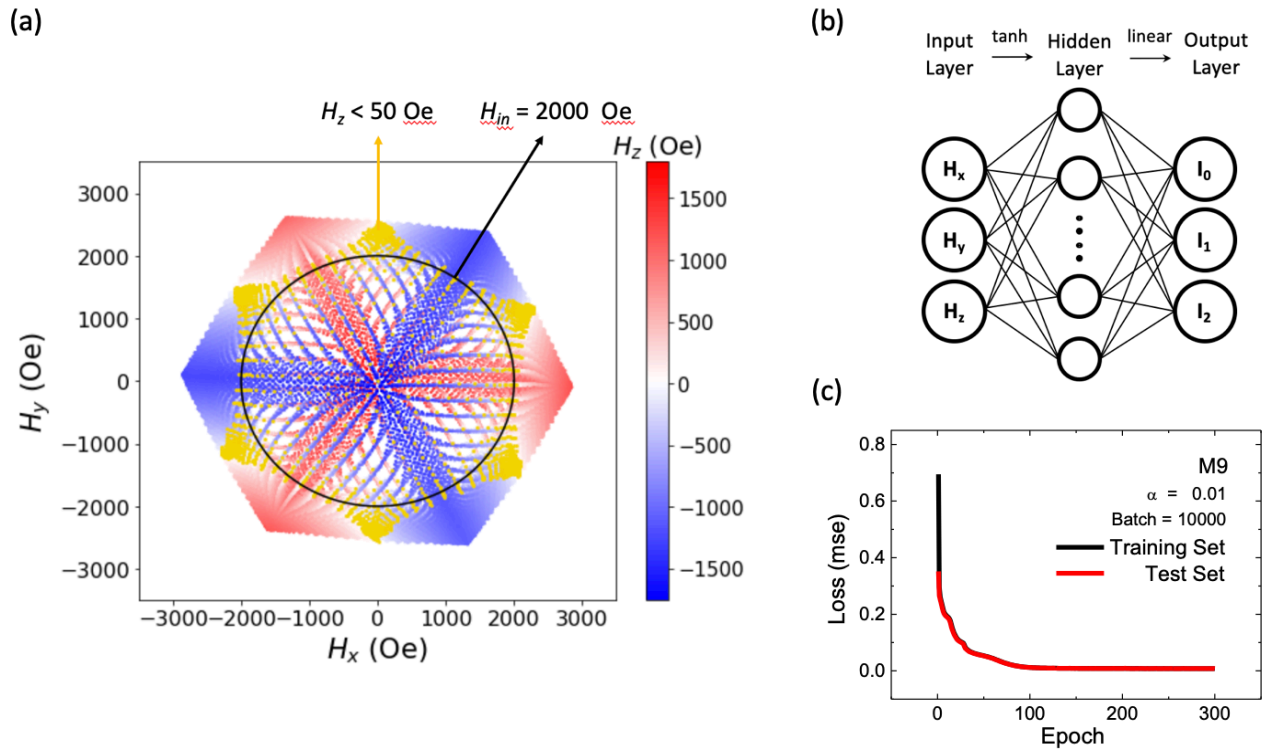

Figure 2. (a) Field map of the projected vector field magnet. The yellow points represent data with out-of-plane magnetic field ($H_z$) less than 50 Oe and the circle represents in-plane field of 2000 Oe. The magnitude of $H_z$ can be referred to the color bar on the right. (b) Schematic plot of the deep learning models with a single hidden layer. (c) Learning curve of model M9 trained with batch size 10,000 and learning rate ($\alpha$) 0.01 for 300 epochs.

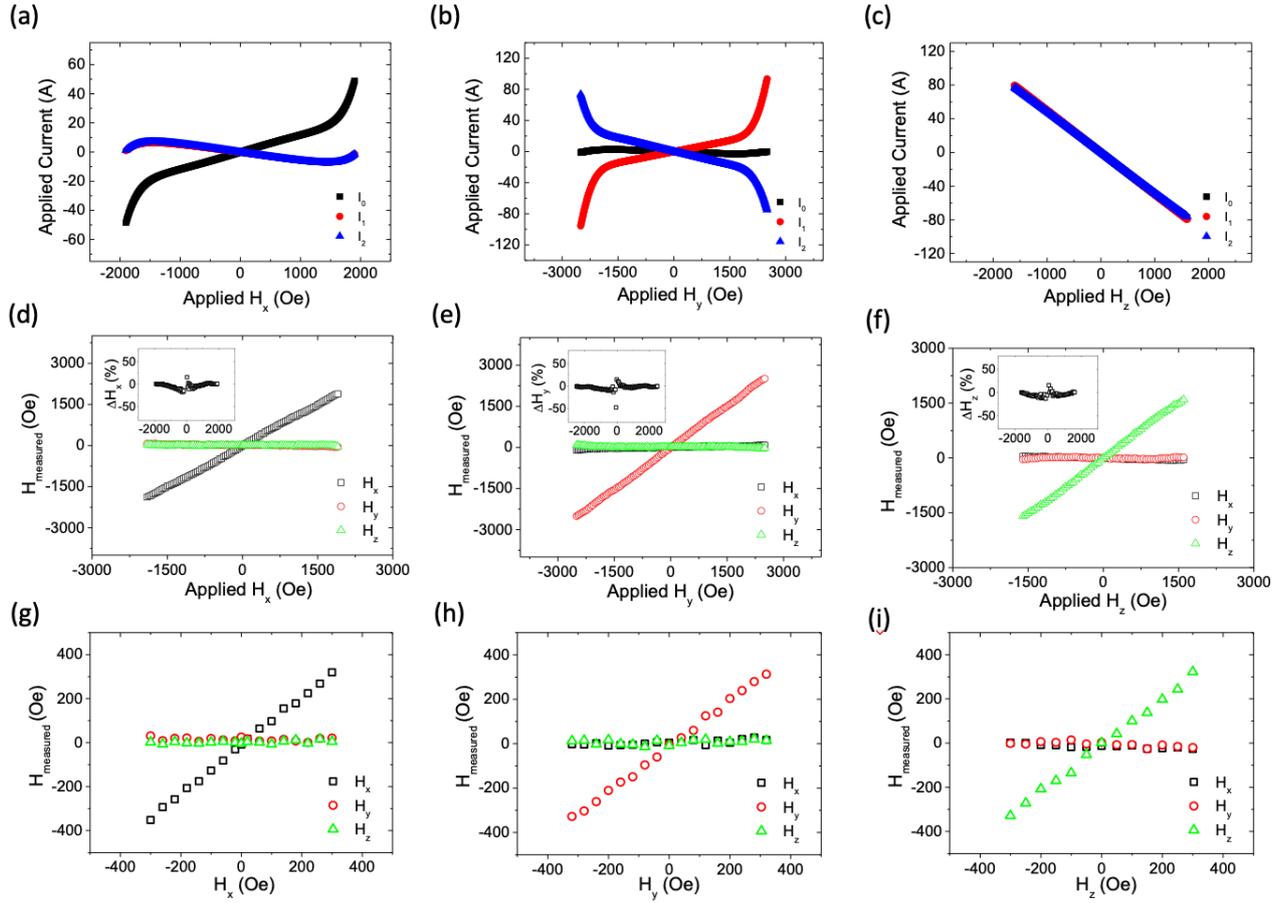

Figure 3. Current predictions and field examinations of field scans. The predicted applied currents as functions of nominal applied field along (a) *x* (b) *y* (c) *z*-axis. The measured fields as functions of nominal applied field along (d) *x* (e) *y* (f) *z*-axis. The relative errors of measured fields as functions of nominal applied field along *x*, *y*, and *z*-axis are the insets of (d), (e), and (f), respectively. The expanded views of measured fields as functions of nominal applied field along (g) *x* (h) *y* (i) *z*-axis.

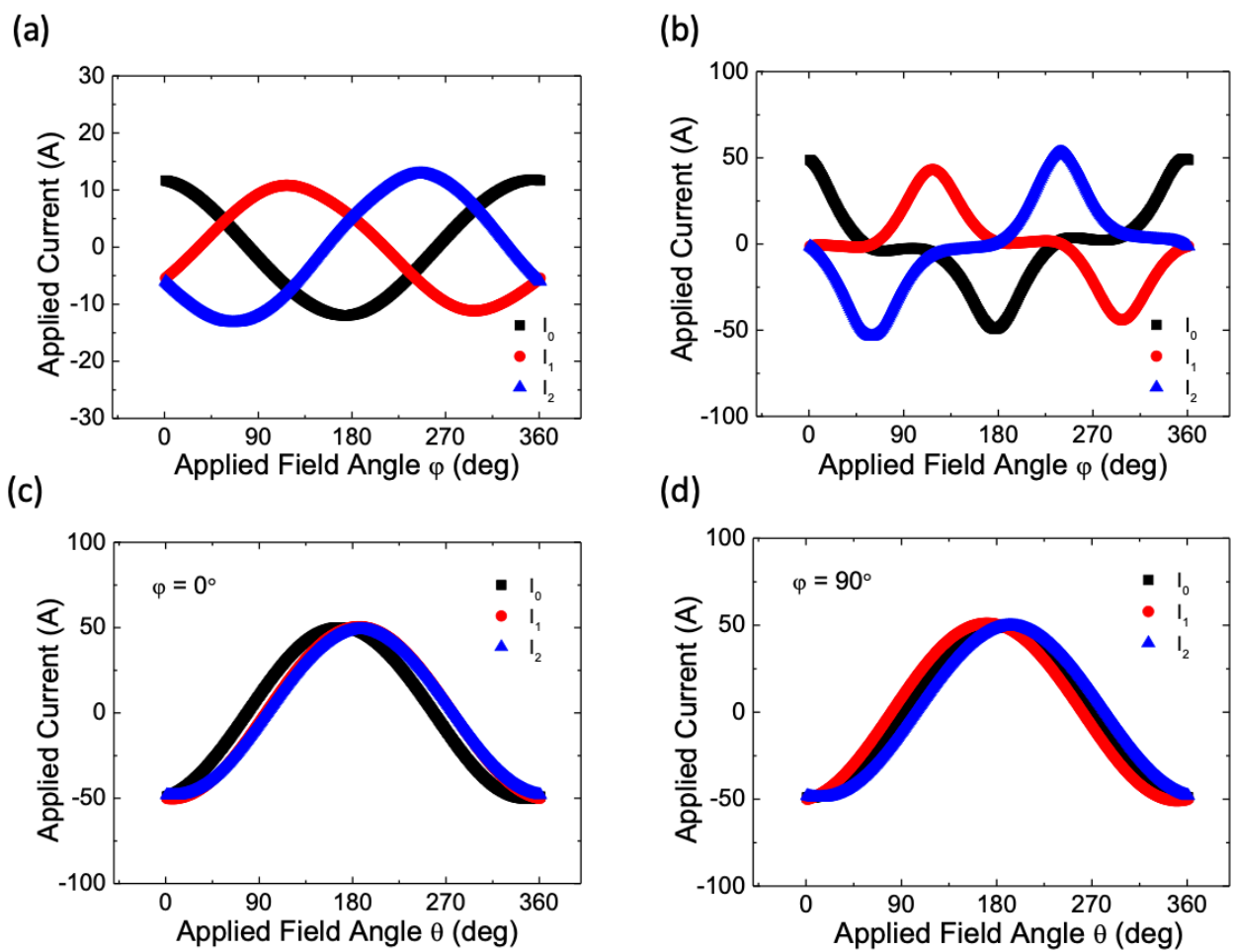

Figure 4. Current predictions of angle scans. The predicted applied currents of $\varphi$ angle scan as functions of $\varphi$ with field magnitude (a) 1000 Oe (b) 1900 Oe. The predicted applied currents of $\theta$ angle scan as functions of $\theta$ with field 1000 Oe at (c) $\varphi = 0°$ (d) $\varphi = 90°$.

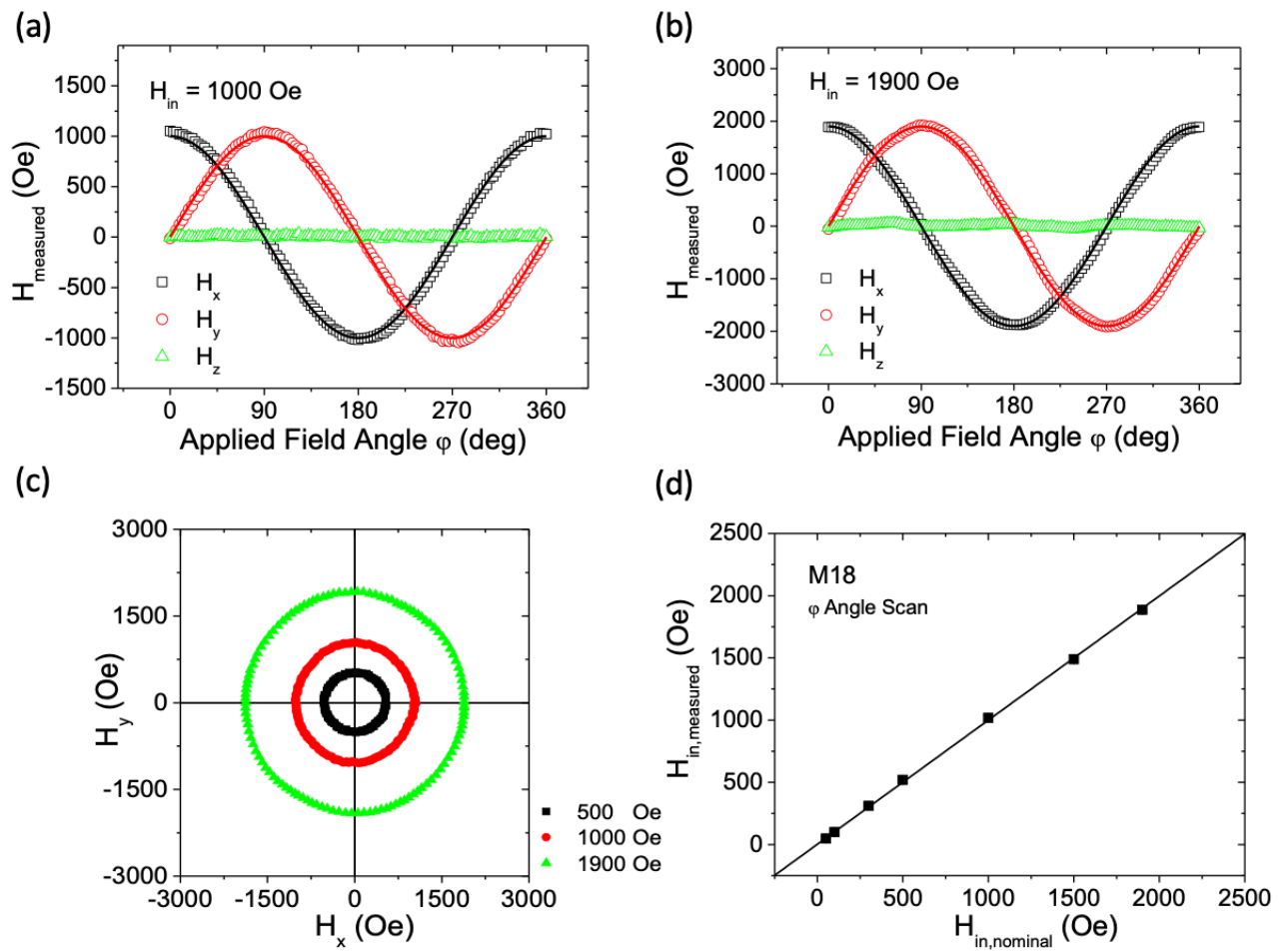

Figure 5. Field examinations of $\varphi$ angle scan. Measured magnetic fields as functions of applied field angle $\varphi$ with field magnitude (a) 1000 Oe and (b) 1900 Oe. (c) Field map of $\varphi$ angle scan at different field magnitudes. (d) The fitted field amplitude as a function of nominal applied field for $\varphi$ angle scan.

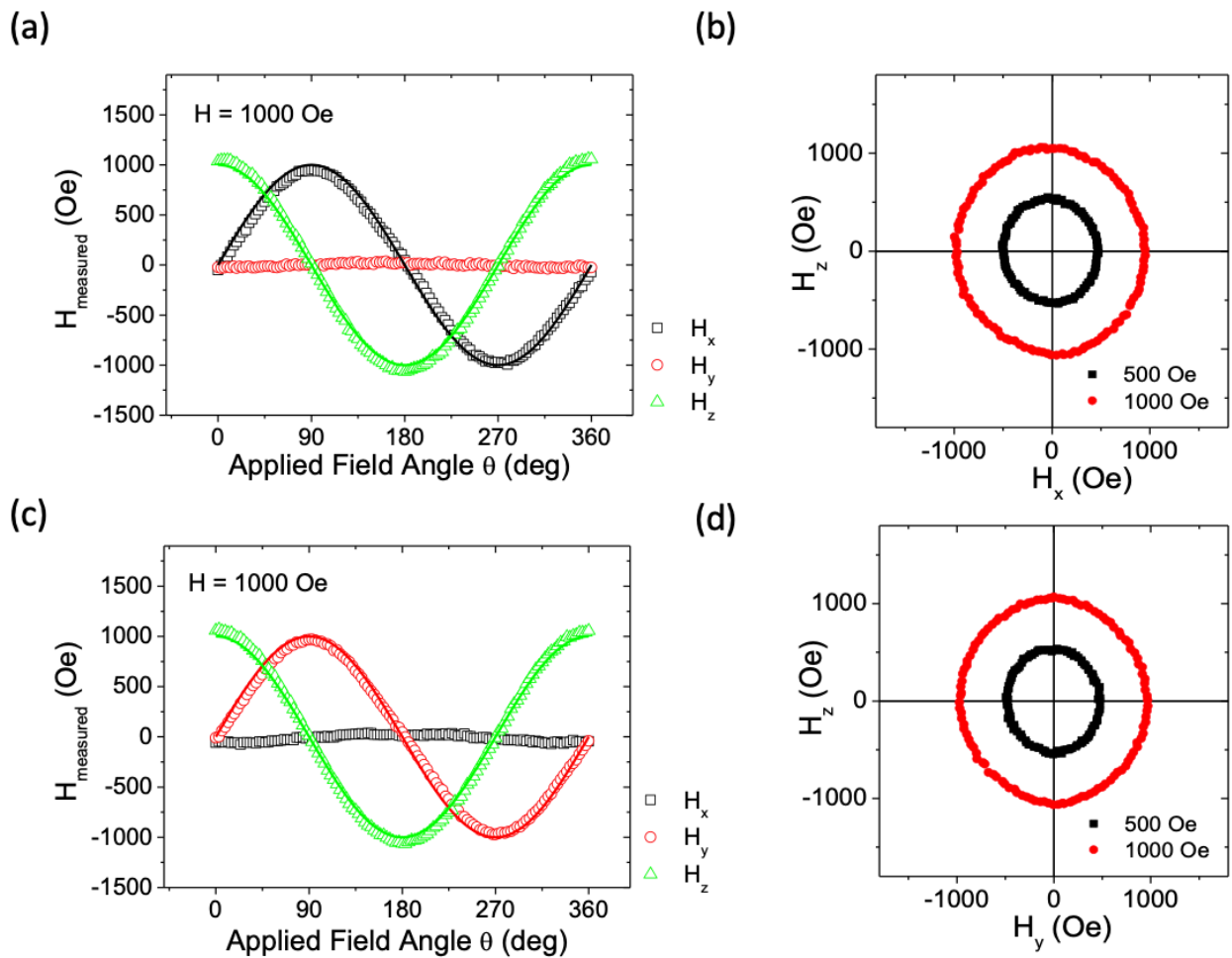

Figure 6. Field examinations of $\theta$ angle scan. (a) Measured magnetic fields as functions of applied field angle $\theta$ on (a) *xz* (c) *yz* plane. Field map of $\theta$ angle scan on (b) *xz* (d) *yz* plane at 500 and 1000 Oe.

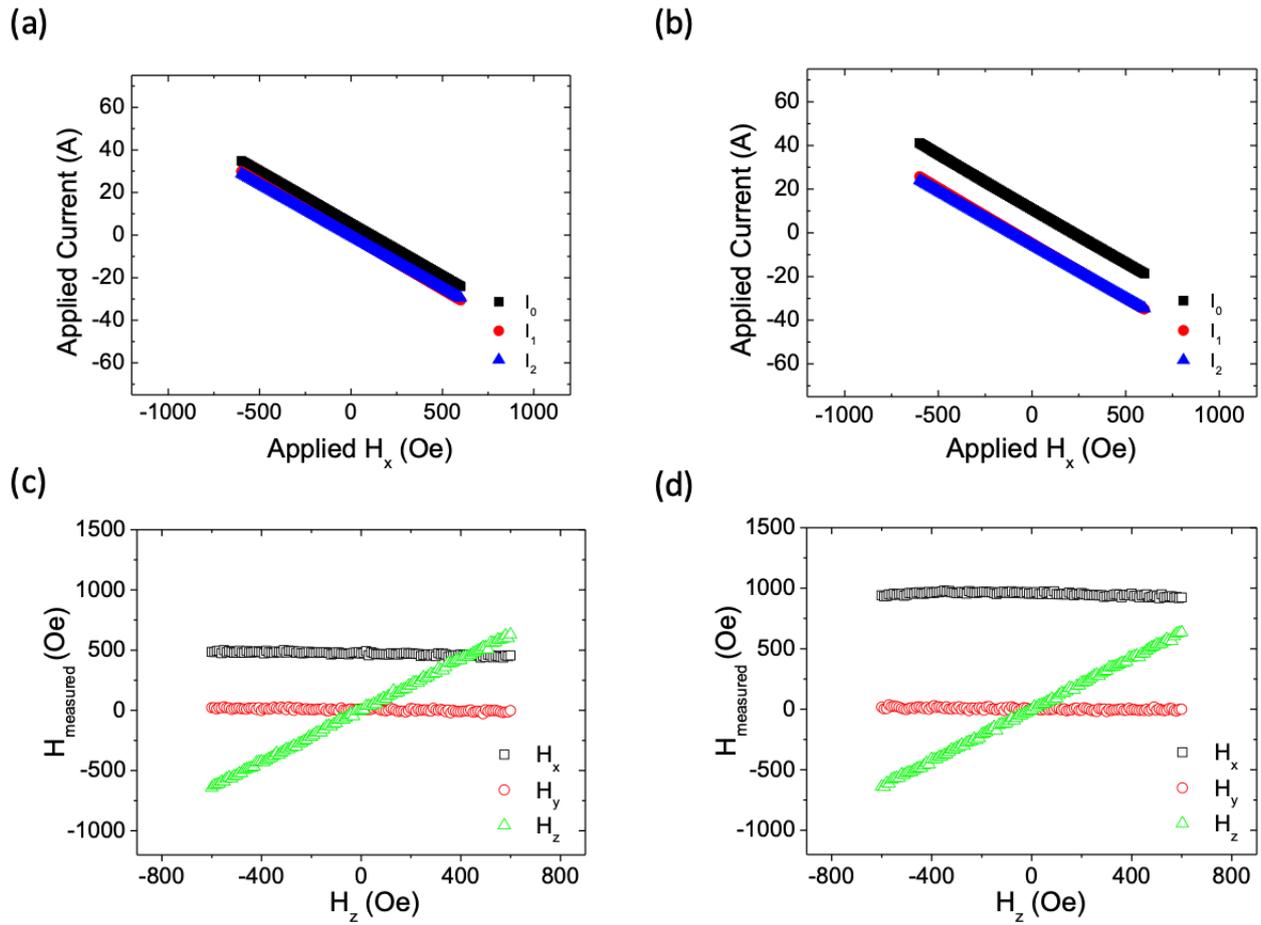

Figure 7. Current predictions and field examinations of hysteresis loop shift measurement. The predicted applied currents as functions of nominal applied $H_z$, where $H_z$ is swept from -600 Oe to 600 Oe with $H_x$ fixed as (a) 500 Oe (b) 1000 Oe. Measured magnetic fields as functions of nominal applied $H_z$ with $H_x$ fixed as (c) 500 Oe (d) 1000 Oe.

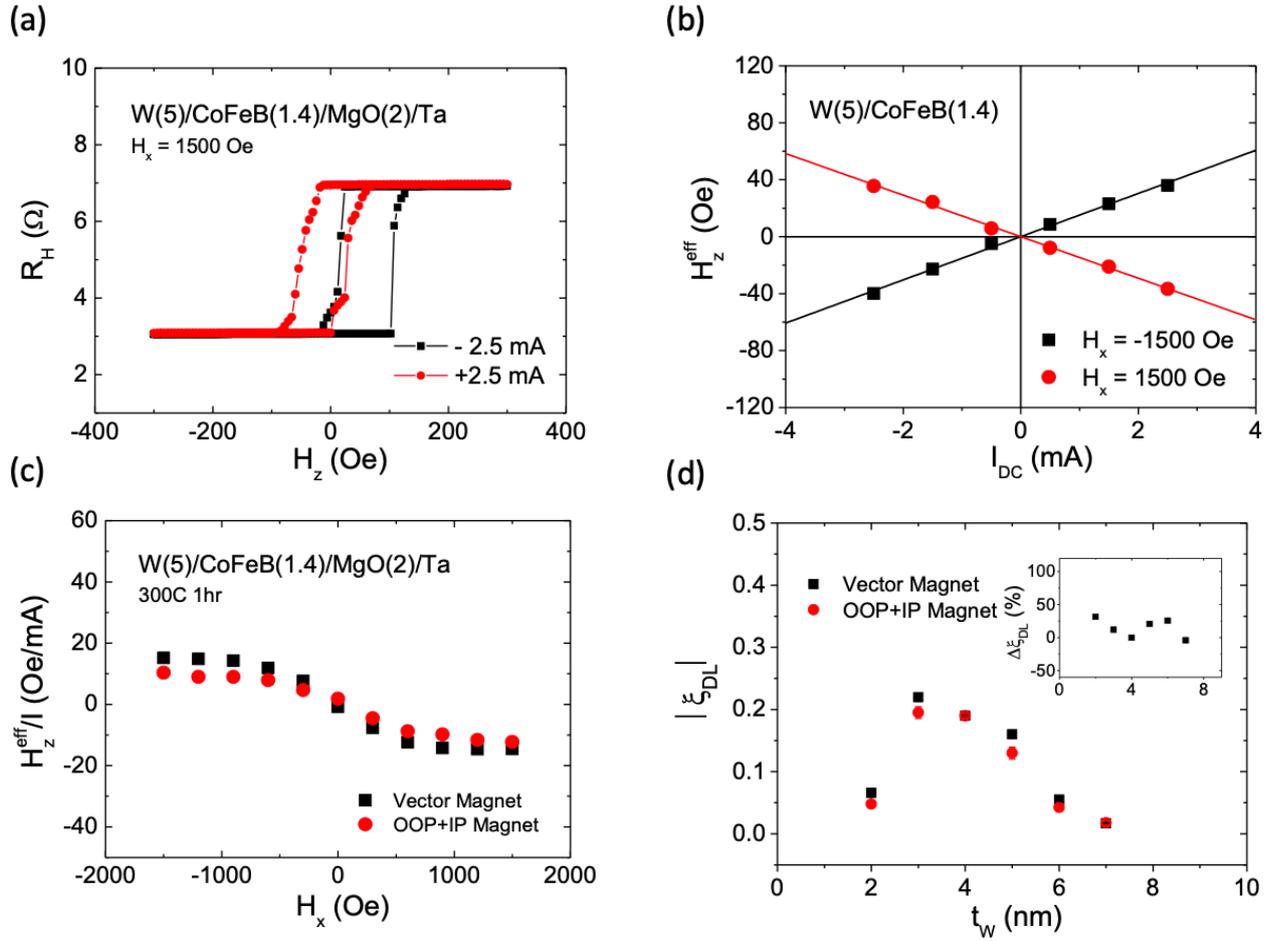

Figure 8. Hysteresis loop shift measurement on a series of W-based Hall bar devices. (a) Representative shifted hysteresis loops of a W(5)/Co-Fe-B(1.4)/MgO(2)/Ta sample with $I_{DC} = \pm 2.5$ mA and $H_x = 1500$ Oe (b) Out-of-plane damping like effective field ($H_{eff}^z$) as a function of applied current under $H_x = \pm 1500$ Oe (c) $H_{eff}^z/I_{DC}$ as a function of applied $H_x$ for W(5)/Co-Fe-B(1.4)/MgO(2)/Ta(2) measured by the projected field vector magnet and a traditional configuration including two separated electromagnets (d) DL-SOT efficiencies as functions of W thickness ($t_W$) extracted with the projected vector field magnet and the traditional setup composed of two separated electromagnets. OOP and IP stand for out-of-plane and in-plane, respectively. Inset is the relative deviation of DL-SOT efficiencies extracted with the two setups.

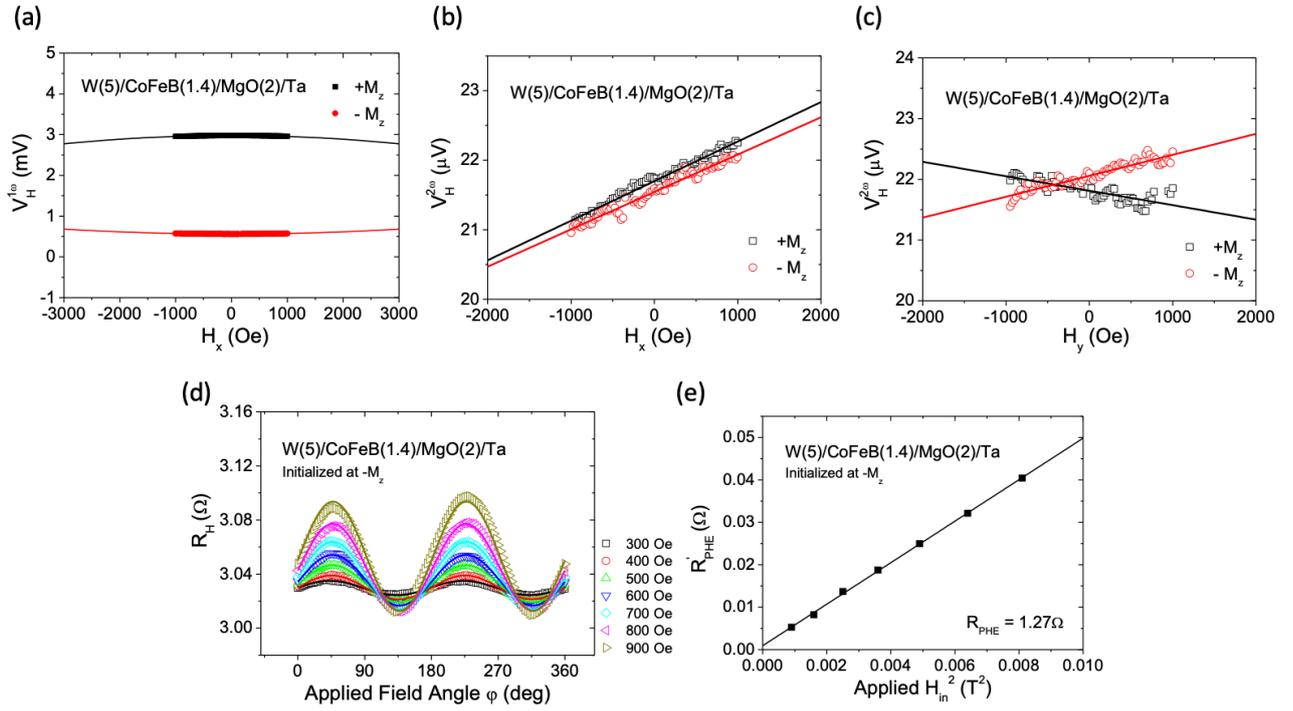

Figure 9. Harmonic Hall measurement on a W(5)/CoFeB(1.4)/MgO(2)/Ta Hall bar device. (a) Representative first harmonic Hall voltage results initialized at $+M_z$ and $-M_z$. The solid curves represent fittings for obtaining curvature and anisotropy field. Representative second harmonic Hall voltages as functions of (b) $H_x$ and (c) $H_y$ with magnetization initialized at $+M_z$ and $-M_z$. (d) Planar Hall resistances as functions of applied field angle with applied in-plane field ranges from 300 to 900 Oe measured after initialization at $-M_z$. The solid curves are the fittings following $\sin 2\varphi$ trend. (e) Fitted $\sin 2\varphi$ amplitude ($R'_{PHE}$) as a function of the square of the applied field.